\documentclass[amsmath, amssymb,aps,prl,twocolumn,superscriptaddress]{revtex4-2}
\usepackage{graphicx}
\usepackage{hyperref}
\usepackage{bm}
\hypersetup{
    colorlinks=true,
    linkcolor=blue,
    filecolor=blue,
    urlcolor=blue,
    citecolor=blue,
}

\begin{document}


\title{
K41 Scaling in Bubble-Induced Turbulence Arises from Single-Bubble Wakes
}

\author{Dabao Li}
\affiliation{School of Engineering Science, University of Chinese Academy of Sciences, Beijing 101408, China}
\author{Zhilong Jin}
\affiliation{School of Engineering Science, University of Chinese Academy of Sciences, Beijing 101408, China}
\author{Guangzhao Zhou}
\email{zgz@ucas.ac.cn}
\affiliation{School of Engineering Science, University of Chinese Academy of Sciences, Beijing 101408, China}
\affiliation{State Key Laboratory of Nonlinear Mechanics, IMECH \& UCAS, Beijing 100190, China}


\begin{abstract}

We use high-resolution, interface-resolved direct numerical simulations (DNS) to investigate the origin of Kolmogorov (K41) scaling in bubble-induced turbulence (BIT).
Region-wise velocity structure functions show that the $2/3$ scaling appears only within bubble wake regions. 
Comparison with matched single-bubble DNS further indicates that the K41 scaling observed in the full BIT field arises from the superposition of individual bubble wakes.
On this basis, we derive a scaling for the dissipation rate in dilute BIT that is consistent with experimental results from the literature.

\end{abstract}


\maketitle


Bubbly flows are ubiquitous in both industrial processes and natural systems~\cite{Magnaudet2000,Mudde2005,Balachandar2010,Takagi2011,Risso2018,Elghobashi2019,Mathai2020,Deike2022}.
In these flows, bubble-induced turbulence (BIT) has attracted sustained attention over the past three decades through numerous experimental and numerical studies~\cite{Lance1991,Bunner2002,Riboux2010,Roghair2011,Prakash2016,Innocenti2021,Ma2022,Pandey2023}.
Most earlier work focused on the regime of pseudo-turbulence, in which velocity fluctuations originate from bubble wakes and the energy spectrum as a function of wavenumber $k$ exhibits a $k^{-3}$ scaling~\cite{MartinezMercado2010,Roghair2011,Prakash2016,Almeras2017,Innocenti2021}.
This regime does not sustain a classical inertial cascade and therefore is not equivalent to fully developed turbulence in the sense of Kolmogorov's 1941 (K41) theory~\cite{Kolmogorov1941}, where a $-5/3$ scaling in the energy spectrum can be observed.
Nevertheless, some observations have hinted at behavior beyond this pseudo-turbulent regime.
In particular, Riboux \textit{et al}.~\cite{Riboux2010} reported a $-5/3$ scaling range at small scales, which they interpreted as probably corresponding to a classical Kolmogorov inertial subrange. More recent studies further identified K41 scaling in BIT at high Galilei number (Ga, the ratio of buoyancy to viscous forces)~\cite{Pandey2023,Ma2025}, suggesting that BIT may transition to a regime consistent with a classical inertial cascade.

DNS by Pandey \textit{et al}.~\cite{Pandey2023} showed that, at sufficiently large Ga, a $-5/3$ scaling emerges in the energy spectrum over $k_d \lesssim k \lesssim 0.3/\eta$, where $k_d \equiv 2\pi/d$, $d$ is the bubble diameter, and $\eta$ is the Kolmogorov length scale.
They further showed that advective energy transfer dominates in this range.
Experiments by Ma \textit{et al}.~\cite{Ma2025} showed that, at a separation distance $r$ smaller than $d$, the longitudinal and transverse second-order structure functions in BIT, $D_{LL}$ and $D_{NN}$, approximately follow K41 scaling, with $D_{LL},D_{NN}\sim r^{2/3}$, corresponding to an energy spectrum $E(k)\sim k^{-5/3}$.
They found that this scaling is outside the bubble wakes, while the in-wake structure functions are steeper than $r^{2/3}$.
They also proposed a dissipation-rate scaling based on the mean bubble spacing.

Although the extent of the identified $k^{-3}$ range differs across studies~\cite{Riboux2010,Innocenti2021}, the reported $k^{-5/3}$ range is comparatively more consistent~\cite{Riboux2010,Pandey2023,Ma2025}.
In particular, the latter scaling is generally observed at scales smaller than $O(d)$.
This suggests that the emergence of K41 scaling in BIT may be closely related to local bubble-scale dynamics.
However, most previous analyses of BIT are based on global statistical averages of the flow field~\cite{Riboux2010,Innocenti2021,Pandey2023}, while physical models have been formulated in terms of collective quantities such as the mean interbubble spacing~\cite{Ma2025}.
This raises a fundamental question: is K41 scaling in BIT governed by a local mechanism associated with individual bubble wakes, or by a collective mechanism arising from many bubbles?

In this Letter, we perform DNS of BIT at relatively high Ga with realistic density and viscosity ratios, fully resolving the bubble wakes.
By separately analyzing wake and non-wake regions, and by comparing the bubble-swarm results with the single-bubble results under the same conditions, we show that the observed K41 scaling in BIT arises from individual bubble wakes.
Based on this finding, we further derive a scaling for the dissipation rate.

The direct numerical simulations are performed with the open-source Basilisk solver~\cite{Popinet2015}.
A volume-of-fluid (VOF) method is used to capture the gas--liquid interface~\cite{Scardovelli1999,Popinet2009}.
The same numerical framework has been validated against benchmark cases for buoyancy-driven bubbly flows and has been used in high-resolution simulations of bubble-induced turbulence (see, e.g., \cite{Innocenti2021}).

All simulations are performed in a cubic periodic domain of side length $L_{\mathrm{box}}$.
The density and viscosity ratios between liquid and gas are fixed at $\rho_l/\rho_g=1000$ and $\mu_l/\mu_g=100$, respectively.
The bubbles are initialized as spheres with the same diameter $d$.
For all cases, $L_{\mathrm{box}}/d$ is approximately ten.
The cases are specified by the gas volume fraction $\alpha$, the Galilei number $\mathrm{Ga}=\rho_l\sqrt{g d^{3}}/\mu_l$, and the E\"otv\"os number $\mathrm{Eo}=\rho_l g d^{2}/\sigma$, where $g$ is the gravitational acceleration and $\sigma$ is the surface tension coefficient.
Unless otherwise specified, all quantities are reported in nondimensional form.
Lengths, velocities, times, and densities are nondimensionalized by
$L_0$, $(gL_0)^{1/2}$, $(L_0/g)^{1/2}$, and $\rho_l$, respectively,
where $L_0=L_{\mathrm{box}}/(2\pi)$.

The parameters of simulated cases are summarized in Table~\ref{tab:cases}.
All simulations are performed using adaptive mesh refinement, with a maximum refinement level of 9, corresponding to a finest-cell spacing equivalent to a uniform resolution of $512^3$.
The results reported in this Letter are insensitive to further grid refinement.
Within the present range of the gas volume fraction, bubble coalescence and breakup are negligible~\cite{Innocenti2021}.
Further details of the numerical method, validation tests, and movies are provided in the Supplemental Material (SM)~\citep{SM}.
\nocite{Chorin1968, Popinet2018, Aniszewski2021, Ramadugu2020, Pope2000, Hidman2023, Xu2025}

\begin{table}[t]
\caption{Parameters and scale separation $d/\eta$ for DNS.
W, NW and L in the last column denote the wake, non-wake, and entire liquid
regions, respectively.
$\eta=(\nu_l^3/\varepsilon)^{1/4}$,
$\nu_l=\mu_l/\rho_l$.
The dissipation rate
$\varepsilon=2\nu_l\langle\boldsymbol{S}\!:\!\boldsymbol{S}\rangle$,
where $\langle\cdot\rangle$ denotes a spatial average over the selected
region, $\boldsymbol{S}=(\nabla\boldsymbol{u}
+\nabla\boldsymbol{u}^{T})/2$, and $\boldsymbol{u}$ is velocity.}
\label{tab:cases}
\setlength{\tabcolsep}{3pt}
\begin{ruledtabular}
\begin{tabular}{cccccc}
Case & $\mathrm{Ga}$ & $\mathrm{Eo}$ & $L_{\mathrm{box}}/d$ & $\alpha$
& $d/\eta\;(\mathrm{W}/\mathrm{NW}/\mathrm{L})$ \\
\hline
1 & 300  & 2.6 & 9.82  & 1.22\% & 34/22/24 \\
2 & 650  & 1.7 & 12.57 & 0.53\% & 59/33/36 \\
3 & 820  & 2.3 & 10.47 & 0.82\% & 71/43/46 \\
4 & 910  & 2.6 & 9.82  & 0.61\% & 74/43/46 \\
5 & 910  & 2.6 & 9.82  & 1.22\% & 76/49/53 \\
6 & 1500 & 2.6 & 9.82  & 1.22\% & 105/69/74 \\
\end{tabular}
\end{ruledtabular}
\end{table}

\begin{figure}
    \centering
    \includegraphics[width=0.8\linewidth]{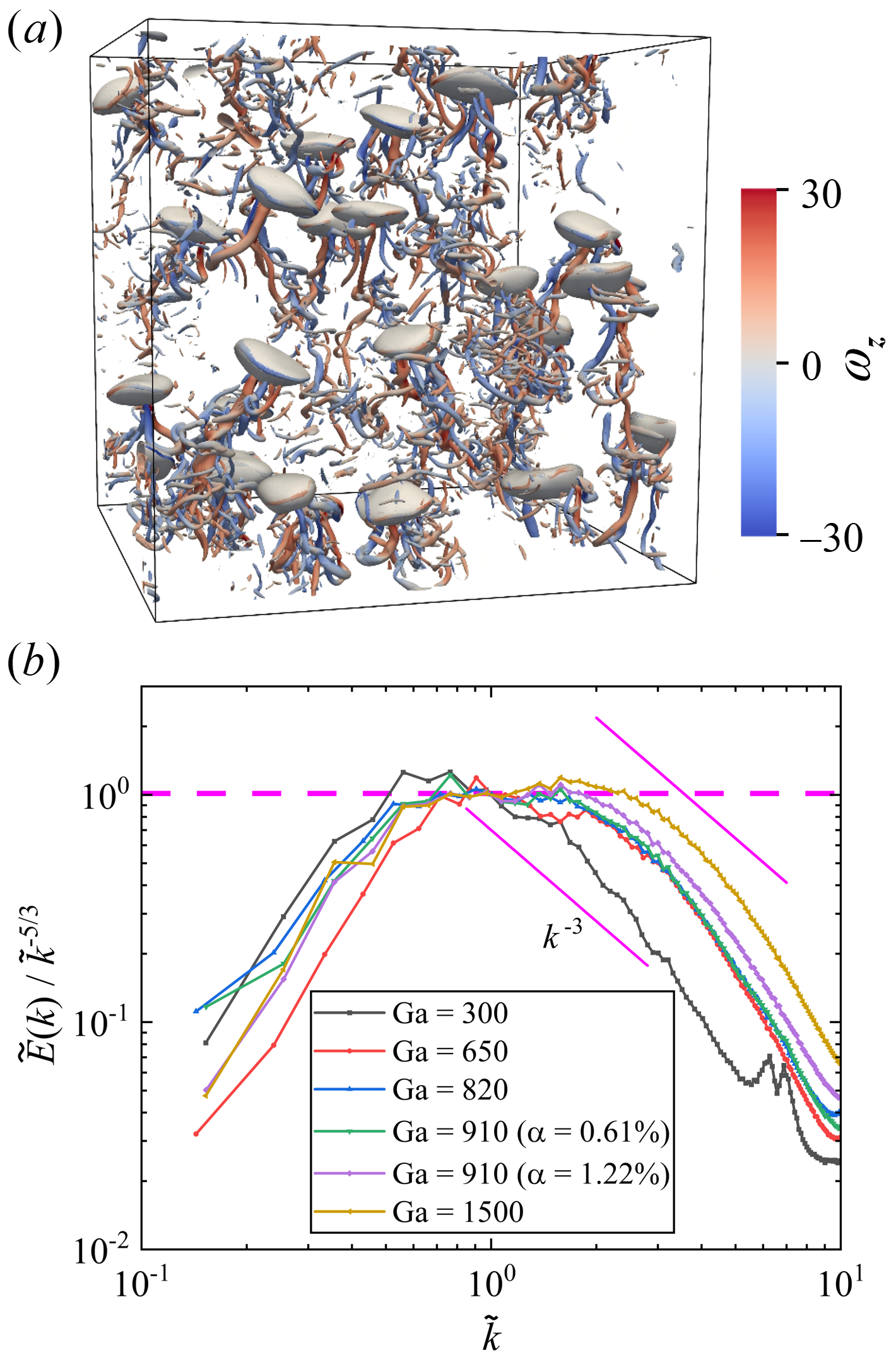}
    \caption{
       (a) Instantaneous vortical structures for $\mathrm{Ga}=1500$,
       visualized using an isosurface of the $Q$ criterion at $Q=50$
       and colored by the vertical vorticity $\omega_z$.
        Here, $Q=(\boldsymbol{\Omega}\!:\!\boldsymbol{\Omega}
          -\boldsymbol{S}\!:\!\boldsymbol{S})/2$, with
          $\boldsymbol{\Omega}
        =(\nabla\boldsymbol{u}-\nabla\boldsymbol{u}^{T})/2$.
       Gray surfaces denote the bubble interfaces.
      (b) Normalized and compensated kinetic energy spectra versus normalized wavenumber. 
      $\tilde{E}=E(k)/E(k_d)$, $\tilde{k}=k/k_d$. 
      The magenta dashed and solid lines indicate the reference $k^{-5/3}$ and $k^{-3}$ scalings, respectively.
}
	\label{fig:overview}
\end{figure}

Figure~\ref{fig:overview}(a) shows the vortical structures identified by the $Q$-criterion at a representative instant in the statistically stationary regime for the $\mathrm{Ga}=1500$ case.
The bubble interfaces and vortical structures in the flow are both well resolved.
The bubbles exhibit significant deformation and deviate markedly from a spherical shape.
As expected, most vortical structures are located beneath the bubbles.

Figure~\ref{fig:overview}(b) shows the kinetic energy spectra for all cases.
The wavenumber is normalized by $k_d$.
The spectrum is normalized by $E(k_d)$ and compensated by $(k/k_d)^{5/3}$, so that the $-5/3$ scaling appears as a plateau as indicated by the magenta dashed line.
No clear $-5/3$ scaling is observed for the low-$\mathrm{Ga}$ cases ($\mathrm{Ga}=300$ and $650$).
By contrast, a $-5/3$ range emerges near $k/k_d \sim 1$ (corresponding to a length scale of $d$) for the higher-$\mathrm{Ga}$ cases.
This range extends towards smaller scales as $\mathrm{Ga}$ increases.
For $\mathrm{Ga}=1500$, the K41 range extends to nearly half a decade, consistent with \cite{Pandey2023}.
The spectrum for $\mathrm{Ga}=300$ is closer to a $-3$ scaling over the range 
$0.7 \lesssim k/k_d \lesssim 2.5$.
For the higher-$\mathrm{Ga}$ cases, however, the $-3$ scaling is not prominent.
It appears only over a short range at scales smaller than the $-5/3$ interval.
Since previous studies have often discussed the $-3$ scaling in connection with scales around the bubble diameter~\cite{Lance1991,Roghair2011,Prakash2016,Innocenti2021}, the present results suggest that, once a $-5/3$ range emerges, it may dominate the spectral interval in which a $-3$ behavior would otherwise be observed.
Consistently, no clear $-3$ scaling was observed in the experiments of \cite{Ma2025}.

\begin{figure}
    \centering
    \includegraphics[width=0.8\linewidth]{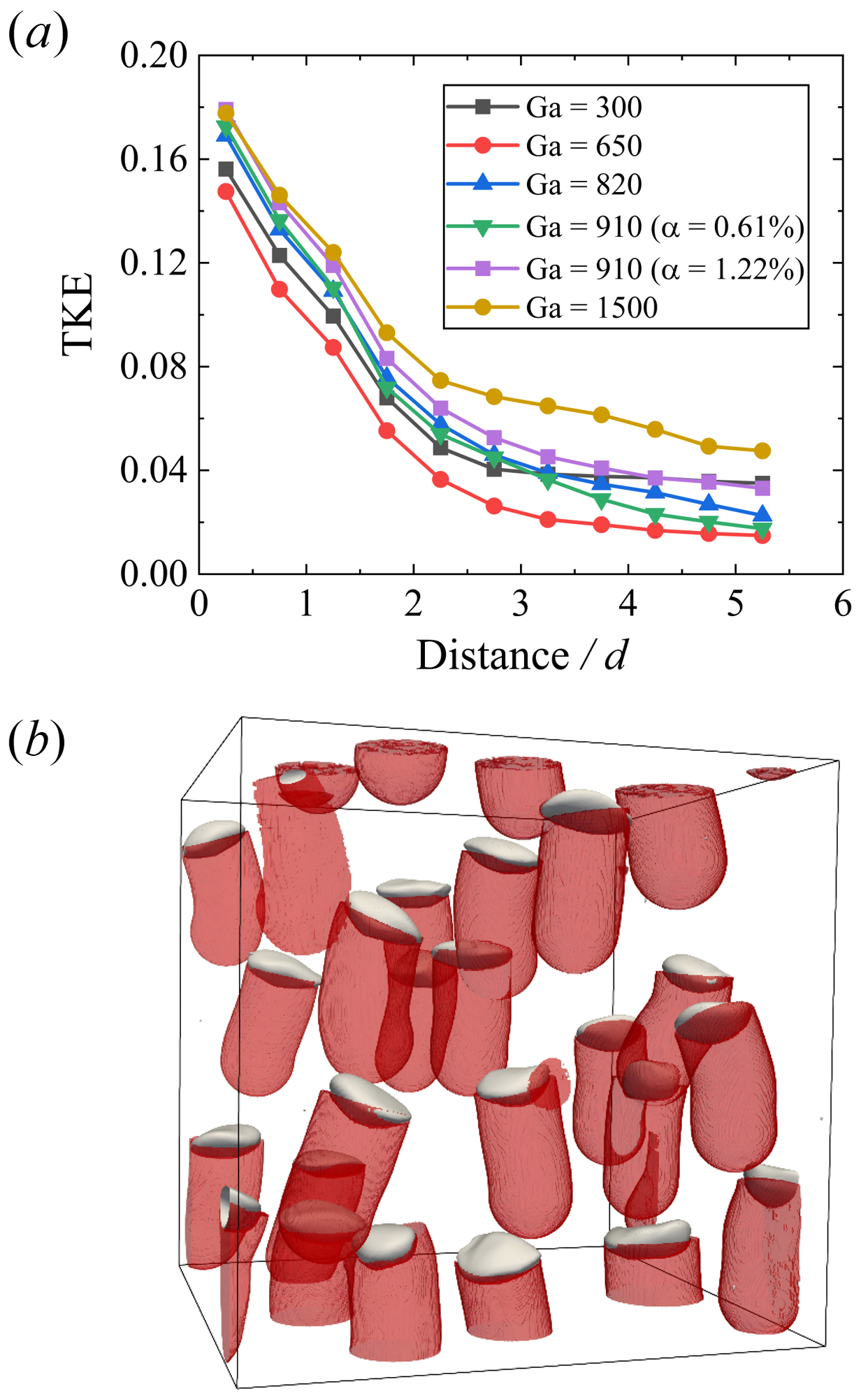}
    \caption{
    (a) Turbulent kinetic energy at the wake centerline, defined as 
    \(\frac{1}{2}\langle u_i'u_i'\rangle\), where 
    \(u_i'=u_i-\langle u_i\rangle\), versus distance along the bubble trajectory.
    (b) Distribution of the identified wake region in the $\mathrm{Ga}=1500$ case, shown at the same instant as in Fig.~\ref{fig:overview}(a). Red volumes denote wakes and gray surfaces denote bubble interfaces.
    }
    \label{fig:wake_length}
\end{figure}

In order to further trace the origin of the K41 scaling, we treat the wake and non-wake regions separately.
Following \cite{Ma2025}, the wake region is identified as the volume swept by a sphere of diameter $d_s$ along the bubble-center trajectory traced backward from its current position over a prescribed length $l$.
The sphere diameter $d_s=1.5d$ is chosen to conservatively account for the effective region over which the deformed bubble strongly disturbs the surrounding liquid.
The wake length $l$ is determined from the decay of turbulent kinetic energy (TKE) along the wake centerline.
As shown in Fig.~\ref{fig:wake_length}(a), the TKE decays rapidly with the distance from the bubble and shows a clear change in slope around $2d$.
This suggests that the wake-induced fluctuations are mainly concentrated within about two bubble diameters behind the bubble.
We therefore set $l=2d$ for all cases.
The identified wake region at the same instant as in Fig.~\ref{fig:overview}(a) is visualized in translucent red in Fig.~\ref{fig:wake_length}(b).
Notably, Ma \textit{et al}.~\cite{Ma2025} adopted a comparatively large wake extent ($d_s=2d$ and $l=5d$) in their work.
The sensitivity of the results to the wake definition is examined
in the SM~\cite{SM}.

\begin{figure}
    \centering
    \includegraphics[width=1.0\linewidth]{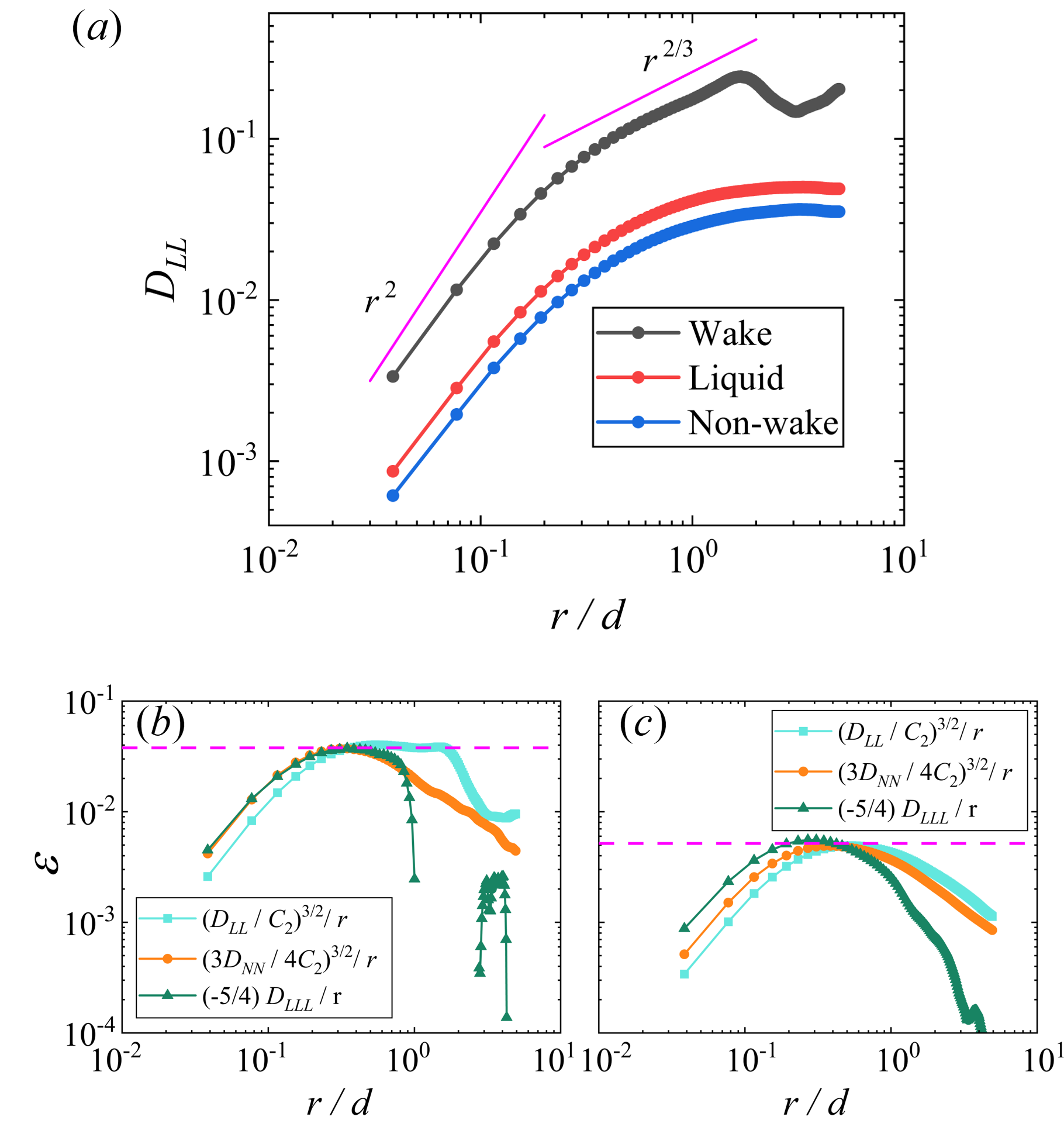}
    \caption{
    (a) $D_{LL}$ as a function of the normalized separation distance $r/d$ for the $\mathrm{Ga}=1500$ case, computed separately in the wake region, the non-wake region, and the entire liquid phase. The two magenta reference lines indicate slopes of $r^2$ and $r^{2/3}$, respectively.
    (b,c) Structure-function-based estimates of the dissipation rate $\varepsilon$ for the $\mathrm{Ga}=1500$ case. 
    The magenta dashed lines indicate the approximate common plateau levels.
    Panels (b) and (c) correspond to the wake region and the entire liquid phase, respectively.
    }
    \label{fig:dll_wake}
\end{figure}

Subsequently, we perform a region-wise analysis of the velocity structure functions, which can be evaluated locally and also provide information complementary to the energy spectrum.
For a separation vector $\boldsymbol{r}$, we define the velocity increment
$\Delta \boldsymbol{u}(\boldsymbol{x},\boldsymbol{r},t)\equiv \boldsymbol{u}(\boldsymbol{x}+\boldsymbol{r},t)-\boldsymbol{u}(\boldsymbol{x},t)$.
Its longitudinal and transverse components with respect to $\boldsymbol{r}$ are denoted by $\Delta u_L$ and $\Delta u_N$, respectively.
The second- and third-order structure functions considered here are defined as
$D_{LL}(r,t)=\langle (\Delta u_L)^2\rangle$,
$D_{NN}(r,t)=\langle (\Delta u_N)^2\rangle$, and
$D_{LLL}(r,t)=\langle (\Delta u_L)^3\rangle$,
where $\langle\cdot\rangle$ denotes a spatial average over the selected region.
Figure~\ref{fig:dll_wake}(a) shows $D_{LL}$ for the $\mathrm{Ga}=1500$ case, computed separately in the wake region, the non-wake region, and the entire liquid phase.
A clear $r^{2/3}$ scaling is observed only in the wake region over $r\in(0.3d,\,2d)$, consistent with the $k^{-5/3}$ scaling range in the energy spectrum.
By contrast, neither the non-wake region nor the entire liquid phase displays a clear scaling behavior.
At sufficiently small scales, all curves tend toward the $r^2$ scaling as is expected in the dissipation range~\cite{Stolovitzky1993}.

To further verify the presence of K41 scaling, we examine the third-order structure function and the corresponding dissipation-rate estimates inferred from the second-order structure functions.
Figures~\ref{fig:dll_wake}(b) and \ref{fig:dll_wake}(c) show the compensated quantities $(D_{LL}/C_2)^{3/2}/r$, $(3D_{NN}/4C_2)^{3/2}/r$, and $-(5/4)D_{LLL}/r$, where $C_2=2.1$ is a constant, for the wake region and the entire liquid phase, respectively.
Under K41 scaling, the three curves are expected to collapse onto a common plateau, with the plateau value equal to the dissipation rate $\varepsilon$~\cite{Kolmogorov1941,Kolmogorov1941b}.
In the wake region, 
such a plateau occurs for $0.2d \lesssim r \lesssim 0.7d$, indicating a K41 cascade over this range.
In the entire liquid phase, the plateau becomes narrower.
This further confirms that K41 behavior is concentrated in the wake region.
The corresponding structure-function results for the other cases are presented in the SM~\citep{SM}.

\begin{figure}
    \centering
    \includegraphics[width=0.75\linewidth]{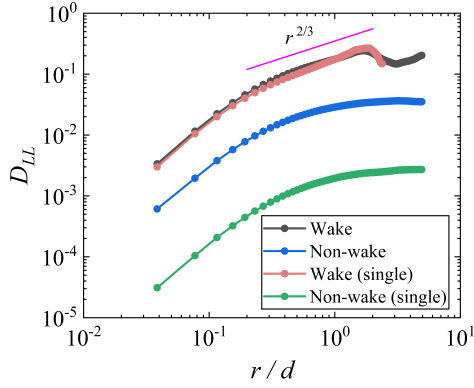}
    \caption{
     $D_{LL}$ for the $\mathrm{Ga}=1500$ case in the wake and non-wake regions of the bubble swarm, compared with the corresponding single-bubble results.
     The magenta reference line indicates the $r^{2/3}$ scaling.
    }
    \label{fig:singlebubble}
\end{figure}

As a further confirmation, we conduct simulations for a single rising bubble with all other parameters identical to the corresponding bubble-swarm cases.
Figure~\ref{fig:singlebubble} compares $D_{LL}$ for the bubble-swarm case and the single-bubble case at $\mathrm{Ga}=1500$.
The structure function in the single-bubble wake (obtained by averaging over five temporally separated wake samples) is found to nearly collapse onto that in the wake region of the bubble swarm.
By contrast, the non-wake structure-function magnitude is much smaller in the single-bubble case than in the bubble swarm.
These results indicate that the K41 scaling observed in the bubble swarm is a reflection of the effects of individual bubble wakes.
Under the relatively dilute conditions considered here, it is expected that the bubble-swarm environment does not alter the individual wake dynamics or the associated cascade behavior, but only enhances the velocity fluctuations outside the wakes.
This explains why the single-bubble parameter $\mathrm{Ga}$ can be used to assess whether K41 scaling can emerge in BIT~\cite{Pandey2023,Ma2025}.
When $\mathrm{Ga}$ is sufficiently large, the corresponding bubble Reynolds number is high enough to create a separation between the scales of energy injection ($\sim d$) and dissipation ($\sim \eta$) within individual bubble wakes (see the $d/\eta$ values in Table~\ref{tab:cases}), such that an inertial cascade can be observed.
Consistently, the two cases at $\mathrm{Ga}=910$, despite a notable
difference in gas volume fraction, exhibit very similar $-5/3$ scaling ranges (Fig.~\ref{fig:overview}(b)).

The above interpretation also motivates a scaling estimate for the kinetic energy dissipation rate $\varepsilon$.
Within the K41 framework, the inertial-subrange energy flux is constant, and scales as $u_r^3/r$~\cite{Vassilicos2015}, where $u_r$ denotes the characteristic velocity of turbulent fluctuations at scale $r$.
This flux is equal to $\varepsilon$.
Therefore, for an individual bubble wake in BIT, we can estimate $\varepsilon_{\mathrm{wake}}$ from the largest inertial-subrange scale by taking $r\sim d$ and $u_r\sim U_b\sim \sqrt{gd}$, where $U_b$ is the single-bubble rise velocity.
This yields $\varepsilon_{\mathrm{wake}} \sim U_b^3/d \sim g^{3/2}d^{1/2}$.
Since dissipation in BIT is mainly concentrated in the wake regions, the global dissipation rate is then calculated by superposition as $\varepsilon \sim \varepsilon_{\mathrm{wake}} V_{\mathrm{wake}}/V$, where $V$ is the total volume and $V_{\mathrm{wake}}$ is the total wake volume.
Since the wake volume associated with each bubble is proportional to the bubble volume, and the overlap between wakes is negligible at low gas volume fraction, one has $V_{\mathrm{wake}}/V \sim \alpha$, yielding
\begin{equation}
\varepsilon = C_{\varepsilon}\,g^{3/2}\alpha d^{1/2},
\label{eq:eps_scaling}
\end{equation}
where $C_{\varepsilon}$ is a dimensionless constant.
Equation~\eqref{eq:eps_scaling} differs from the scaling in Ref.~\cite{Ma2025}, which treats BIT as a collective flow and takes the interbubble spacing, $d\alpha^{-1/3}$, as the characteristic length scale in the dissipation estimate.

\begin{figure}[h]
    \centering
    \includegraphics[width=0.75\linewidth]{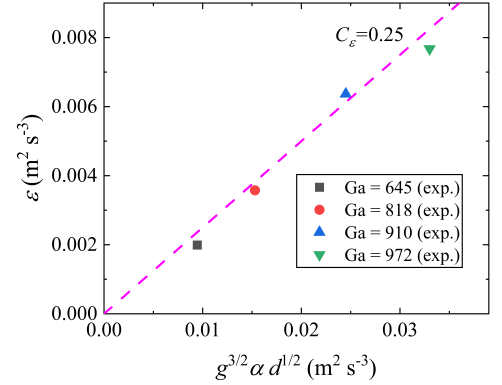}
    \caption{
     Kinetic energy dissipation rate $\varepsilon$ as a function of $g^{3/2}\alpha d^{1/2}$, using the experimental data extracted from Fig.~4 of \cite{Ma2025}. 
     In \cite{Ma2025}, $\varepsilon$ was estimated from compensated structure functions obtained from 3D Lagrangian particle tracking measurements.
     The dashed line represents Eq.~\eqref{eq:eps_scaling} with $C_\varepsilon=0.25$.
     The variables in this figure are dimensional.
    }
    \label{fig:dissipation}
\end{figure}

To assess this scaling, Fig.~\ref{fig:dissipation} plots $\varepsilon$ against $g^{3/2}\alpha d^{1/2}$ using the experimental data extracted from Fig.~4 of \cite{Ma2025}.
The data are broadly consistent with a linear dependence.
Moreover, they are reasonably described by a line passing through the origin, consistent with the physically trivial limit that $\varepsilon \to 0$ when the driving vanishes, for example as $\alpha \to 0$ or $g \to 0$.

It is worth noting that Lance and Bataille~\cite{Lance1991} estimated the BIT dissipation rate from the power input associated with bubble drag, leading to the scaling $C_D \alpha U_b^3 / d$, where $C_D$ is the drag coefficient.
Pandey \textit{et al}.~\cite{Pandey2020} further rewrote this expression as $C_D g^{3/2}\alpha d^{1/2}$, whose form is identical to that of Eq.~\eqref{eq:eps_scaling}.
This agreement is physically expected, since under statistically stationary conditions, the energy input, the inertial-subrange energy flux, and the final dissipation rate are equal on average within the K41 framework.

In summary, our high-resolution DNS show that the $r^{2/3}$ scaling is confined to bubble wake regions. 
The K41 scaling observed in the entire BIT field originates from inertial cascades within individual bubble wakes and can be interpreted as the superposition of contributions from these wakes.
A wake-based scaling law for the dissipation rate in dilute BIT is proposed, which agrees well with existing experimental data and theoretical work.
These results provide new insight into the origin of Kolmogorov scaling in bubble-induced turbulence.

\vspace{1em}
\noindent \textit{Acknowledgements}
The authors acknowledge the support of the National Natural Science Foundation of China (Grant
No. 12572298).

\bibliography{apssamp}

\appendix

\end{document}